\documentstyle[12pt]{article}

\title{\bf     BINARY SYSTEMS IN QM AND IN QFT: CPT}
\author{Victor Novikov \\
\it ITEP, Moscow, Russia}
\date{}

\def\fun#1#2{\lower3.6pt\vbox{\baselineskip0pt\lineskip.9pt
\ialign{$\mathsurround=0pt#1\hfil##\hfil$\crcr#2\crcr\sim\crcr}}}

\begin{document}

\maketitle

\begin{abstract}

Quasi-degenerate  neutral systems like a $(K, \bar{K})$
type are investigated in Quantum Field Theory (QFT). A constant mass matrix as the one used in Quantum Mechanics (QM)
can only be introduced as a linear approximation to QFT.
We study the phenomenological consequences of the differences between the QFT
and QM treatments.
The role of ``spurious'' states with zero norm at the poles  is emphasized.
The $K_L-K_S$ mass splitting triggers a tiny difference between the  $CP$
violating parameters $\epsilon_L$ and $\epsilon_S$, without any
violation of $TCP$.
Non-vanishing  semi-leptonic asymmetry
$\delta_S - \delta_L$  does not signal $TCP$
violation (usual claims not withstanding), while $A_{TCP}$ keeps vanishing when $TCP$ is good symmetry.

\end{abstract}

\newpage

This talk is based on the paper written  in
collaboration with B. Machet and M.Vysotsky\cite{0}.

\section{Introduction}

I would like to reconsider the theory of  binary quasi-degenerate neutral
systems such as $(K,\bar{K})$, $(D,\bar{D})$ and $(B,\bar{B})$.
The result of this revision will be the well-known conventional theory plus tiny corrections.
It seems  unlikely that these corrections can be ever measured experimentally
but conceptually they are rather interesting.

 The  $(K,\bar{K})$ meson system is among the most magnificent ones in particle physics.
Fifty years ago  Gell-Mann, Pais, and Pais and Piccioni\cite {Pais}
discovered and described in great detail such beautiful phenomenon as K meson
oscillations. Later  $CP$ symmetry violation \cite{ChristensenCroninFitchTurlay}  was
found in  $(K,\bar{K})$ system. Nowadays the search for   $CP$ violation in   $(B,\bar{B})$ system is a  hot topic in
experimental particle physics. Binary systems are also  well
 suited  to test   $CPT$ symmetry  and Quantum Mechanics ($QM$) in general.

The main property of the binary systems is that the
 splitting between two states in binary systems is extremely small in comparison with other
masses. Thus it is widely believed that with good accuracy one can separate
the dynamics of quasi-degenerate states from the details of the dynamics of other
states. In other words one can  integrate over infinite number of degrees of freedom
in Quantum Field Theory ($QFT$) and deal with  the rest finite number of degrees of freedom, i.e. with
 effective  Quantum Mechanics.

According to this philosophy in the case of binary systems one deals with effective $QM$ with
 $2\times2$ non-hermitian  Hamiltonian

\begin{equation}
H =M- \frac{i}{2}  \Gamma,
\end{equation}
where  $M=M^{\dagger},\;  \Gamma=\Gamma^{\dagger} $.
The details of $QM$ treatment of  binary systems  can be found in text-books. A brief and a
very transparent version of this conventional theory is presented in  Review of Particle
Physics \cite{RPP}. As a rule the text-books do not
go beyond ($QM$)  level in treatment of $(K,\bar{K})$ system. Only recently the need of a
treatment of these systems in the
framework of Quantum Field Theory  arose \cite{BlasoneCapolupoRomeiVitiello},
\cite{Beuthe}. It was actually mainly motivated by the
leptonic sector, i.e. by the attempts to treat  neutrino oscillations in terms of $QFT$.

The formalism of a mass matrix  seems never to be in doubt, though
its existence, as we shall see, can only be assumed in a certain
approximation.
The growing need for precise criteria to test
discrete symmetries made necessary an exhaustive investigation of these
systems in $QFT$.  This was done in \cite{0} and a short version of this
study is  presented here.

\section{ $K\bar{K}$ system in $QM$ }

First let me remind the conventional $QM$ wisdom for $(K, \bar{K})$ system.
In  Wigner-Weisskopf approximation
to describe decaying particles we use non-hermitian effective Hamiltonian

\begin{equation}
H =M- \frac{i}{2}  \Gamma,
\end{equation}
with mass matrix  $M=M^{\dagger} $  and decay matrix  $ \Gamma=\Gamma^{\dagger}$ . Thus

\begin{equation}
{H} = \left( \begin{array}{cc}
   m_{11} -\frac{i}{2} \gamma_{11}  &  m_{12} -\frac{i}{2} \gamma_{12}\cr
  \overline{m_{12}} -\frac{i}{2}\overline{\gamma_{12}} & m_{22}-\frac{i}{2}\gamma_{22}
\end{array}\right),\
\end{equation}
where $ m_{11}, m_{22},\gamma_{11},\gamma_{22}$ are real numbers. Matrix elements of $M$ and $\Gamma$
can be considered as a set of phenomenological parameters that can be extracted from the experimental data.
On the other hand the same matrix elements  $H_{ik}$ of the effective Hamiltonian can be connected with matrix
elements of the fundamental weak Hamiltonian $H_W$.
For example

\begin{equation}
 M_{ik}=m\delta_{ik} + <i| H_W |k> + \sum_{\beta} <i|H_W|{\beta}><{\beta}|H_W|k>/(m-E_{\beta}),
\end{equation}
where $i, k$  numerate $K^0$ $\bar{K^0}$ mesons and $\beta$ numerates all other states.

The eigenvectors that correspond to ''stationary'' non-mixing states are  $K_S$  and $K_L$  states

\begin{equation}
|K_S>=\sqrt{N_S} (\left |\ K^0_1> +\epsilon_{S}|\ K^0_2>\right),\;\;
|K_L>=\sqrt{N_L} (\left |\ K^0_2> +\epsilon_{L}|\ K^0_1>\right),
\end{equation}
where $|\ K^0_1>$ and $|\ K^0_2>$ are $CP$-even and $CP$-odd states.
Mixing parameters $\epsilon _{S, L}$ can be calculated in terms of matrix elements $H_{ik}$ of effective Hamiltonian.

There are few constraints on the elements $H_{ik}$  that follows from the general symmetries of the system.

Thus from CPT symmetry one can derive that diagonal elements of effective hamiltonian are identical, i.e. $H_{11} = H_{22}$.
From this equation one gets that mixing parameters  $\epsilon _{S, L}$ are
 the same    $\epsilon_S = \epsilon_L$.

For  CP-symmetric interaction one derives additional constraint on
the non-diagonal elements. Namely, one gets that
$H_{12} =e^{i \alpha} H_{21}$.
 From this equation  it follows that mixing parameter is zero  $\epsilon = 0 $,
 and that eigenvectors of effective Hamiltonian are the states with definite CP-parity, i.e. $|\ K^0_1>$ and $|\ K^0_2>$.

One can calculate all $CP$ and $CPT$ violating
asymmetries in terms of mixing parameters $\epsilon_{L,S}$.

Formalism of Mass Matrix for  $K\bar{K}$ system seems never to be in doubt, though in the very early publications people
mentioned possible corrections to Wigner- Weisskopf approximation. (Although see ref. \cite {Azimov}).
\section{Normal and non-normal Quantum Mechanics}

 Quantum Mechanics with non-hermitian Hamiltonian eq. (2) is rather different from the conventional $QM$.
The reason is that in general case the effective Hamiltonian eq.(2) is not a normal operator.

 Let me recall a definition of a normal matrix:

\begin{equation}
M \; normal \Leftrightarrow [M,M^{\dagger}]=0.
\end{equation}

Normality is a remarkable property of matrices:

$1)\;$any matrix that commutes with its hermitian conjugate can be diagonalized
by a single unitary transformation;

$2)\;$its right and left eigenstates accordingly coincide;

$3)\;$it admits complex eigenvalues, which makes it specially suited to describe unstable particles.

When $CP$ is conserved, we have shown that the propagator of neutral kaons
and effective Hamiltonian $H$
must be normal. This will provides us with the most general $CP$
eigenstates in the $(K^0,\overline{K^0})$ basis.

It is very tempting to have a normal propagator and normal effective Quantum Mechanics,
since in this case the right eigenstates and left eigenstates coincide. Unfortunately this is
impossible.
In general the mass matrix $M$ does not commute with the decay matrix $\Gamma$,

\begin{equation}
[M, \Gamma]\not=0.
\end{equation}
Thus Hamiltonian $H$ does not commute with its hermitian conjugate  $H^{\dagger}$

\begin{equation}
[H, H^{\dagger}]\not=0,
\end{equation}
and is non-normal operator.

In this case  the left and right eigenstates

\begin{equation}
H \;|in> = E_{in}\; |in>, \;\; <out|\; H =  <out|\; E_{out}
\end{equation}
are independent sets of vectors, i.e. they are not connected by complex conjugation

\begin{equation}
<out| \not=  |in>^{\dagger}.
\end{equation}
As for the eigenvalues, one can prove that
\begin{equation}
E_{out} = E_ {in}.
\end{equation}

For $K$-meson system all these can be rewritten as
\begin{equation}
H \;|K_{L,S}> = (m_{L,S}-\frac{i}{2} \gamma_{L,S})\; |K_{L,S}>,
\end{equation}
 \begin{equation}
<K_{L,S}|\;H = (m_{L,S}-\frac{i}{2} \gamma_{L,S})\; <K_{L,S}|
\end {equation}

Left eigenstates are orthogonal to the right eigenstates
 \begin{equation}
  <K_{L,S}|K_{S,L}>=0
 \end{equation}
and there is no complex conjugation, i.e.
 \begin{equation}
 <K_{L}|\not= |K_{L}>^{\dagger}.
 \end{equation}
  All these mean  that $<K_{L}|_{out}$ and $|K_{L}>_{in}$ are
different mixtures  of $K$ and $\bar{K}$ state. This statement, being formally
absolutely correct, is rather unconventional and encourages us to look for different description of binary systems.

\section{Effective QFT approach}

It seems natural to start from  $QFT$ and to derive effective $QM$ as some approximation to $QFT$.
Partly this line of reasoning was motivated by  numerous attempts to develop theory of
$\nu$-oscillation in terms of Green functions.

We work within effective Field Theory where $K$-mesons, pions, etc are considered as
elementary particles that are described by the corresponding field operators
$\phi_{K}(x)$, $\phi_{\pi}(x)$, etc.
The propagator of these particle are given by  v.e.v. of T-product of the appropriate field operators.
Say propagation of $K^0$ into $K^0$ is equal to

\begin{equation}
 <K^0|\Delta(x)|K^0> = <0|\ T \{ \phi_{K^0}(x), \; \phi^{\dagger}_{K^0}(0)\}\;|0>.
\end{equation}

For $(K,\bar{K})$ mesons quasi-degenerate system the propagator  is described by  $2\times2$ matrix

\begin{equation}
{\Delta}(z) = \left(\begin{array}{rr}
                <K|\Delta|K> & <K|\Delta|\bar{K}> \cr
 <\bar{K}|\Delta|K > & <\bar{K}|\Delta|\bar{K}>\end{array}\right)\,
\end{equation}
where $z=q^2$  and $q$ is momentum.
For any momenta $q$ one can diagonalize $\Delta$  and find corresponding eigenstates, i.e.

\begin{equation}
\Delta(z) |R_{\pm}(z)> = \lambda_{\pm}(z) |R_{\pm}(z)>,
\break
<L_{\pm}(z)|\Delta(z) =  <L_{\pm}(z)|\lambda_{\pm}(z).
\end{equation}
The eigenstates are ``stationary'' states ( i.e. there is no oscillation between $K_L$ and $K_S$).
Eigenvalues of the propagator  $\lambda_{\pm}$ are the same for $in$ and $out$ states.
There is no reason for $\Delta(z)$ to be normal. Thus complex
conjugation does not transform left states into right states and visa verse, i.e.
\begin{equation}
<R_{\pm}|\equiv |R_{\pm}>^{\dagger} \Leftrightarrow <R_{+}| R_{-}>\not= 0.
\end{equation}
This is in one-to-one correspondence with $QM$ approach.

One can write Dyson-Schwinger equation for all 4 propagators. For inverse matrix $\Delta^{-1}(z)$ it looks like

\begin{equation}
{\Delta^{-1}} = \left( \begin{array}{cc}
   q^2-m^2-\Pi_{KK}(q^2) &\Pi_{K \bar{K}}(q^2)\cr
   \Pi_{\bar{K} K}(q^2) &    q^2-m^2-\Pi_{\bar{K}\bar{K}}(q^2)
\end{array}\right),\
\end{equation}
where $(\Pi_{KK}(q^2),\Pi_{\bar{K}\bar{K}}(q^2))$ and
$(\Pi_{K\bar{K}}(q^2),\Pi_{\bar{K}K}(q^2))$ are diagonal and
non-diagonal self-energy functions.  Dyson-Schwinger equations
for these self-energy functions include vertex operators. There are infinite number of equations for vertex operators and
$QFT$ with its infinite number of degrees of freedom exhibits itself exactly at this level. But whenever
self-energy functions are known  one can describe $(K^0, \bar{K}^0)$
system in terms of $2 \times 2$ propagators matrix. These functions are analog of matrix elements $H_{ik}$ in $QM$ approach. 

Actually to construct a  bridge between $QFT$ and $QM$ we need to know a little about these
self-energy functions.
To proceed it is useful to consider analytical properties of propagator.

\subsection{K\"allen-Lehmann representation}

{\underline {\bf Analyticity:}}
It  can be demonstrated, with very general hypothesis that  propagator satisfies a K\"allen-Lehmann representation \cite{SW}, which is written, in Fourier space, as

\begin{equation}
\Delta(z) = \int_0^\infty d k^2 \frac{\rho(k^2)}{k^2-z},\
\end{equation}
where, eventually, $z$  gets close to the cut on the real axis by staying in
the physical upper half-plane $z \rightarrow (p^2 + i\varepsilon),  p^2 \in
{\it R}$. A consequence is that the propagator $\Delta(z)$  is an holomorphic function in the complex $z$ plane outside the cuts.
\break{{\underline {\bf Positivity:}}}
The spectral function $\rho (k^2)$ is  a positive hermitian matrix.
A consequence is that the propagator $\Delta(z)$  in the complex $z$ plane outside the cuts satisfies \cite{SW}

\begin{equation}
\Delta(z) = [\Delta(\bar z)]^\dag.
\end{equation}

Indeed, one can write, using the hermiticity of $\rho$

\begin{equation}
\Delta(\bar z) = \int_0^\infty dk^2 \frac{\rho (k^2)}{k^2-\bar z},
\break
[\Delta(\bar z)]^\dag = \int_0^\infty dk^2
\left[\frac{\rho(k^2)}{k^2-\bar z}\right]^\dagger =
\int_0^\infty dk^2 \frac{\rho^\dagger(k^2)}{k^2-z}.\
\end{equation}

This general property should be distinguished from the (Schwarz)
reflection principle or its refined version called the ``edge of the wedge''
theorem \cite{SW}; indeed, as soon as complex coupling
constants can enter the game, in particular to account for $CP$
violation, the discontinuity on the cut is no longer the sole origin for
the imaginary part of the propagator; it can be non-vanishing outside the
cut, which is likely to invalidate
the principle of reflection.

In $QFT$, the physical masses are the poles of propagator $\Delta(z)$, i.e.

\begin{equation}
Mass\;\; states  \Leftrightarrow Det (\Delta^{-1}(q^2))=0.
\end{equation}

Thus for binary system of $K$-mesons we have two complex poles

 \begin{equation}
z_{1} = M_{L}^{2}, \;\; z_{2} = M_{S}^{2}.
\end{equation}

\subsection{Introducing a mass matrix: from QFT to QM}

Here we demonstrate how the effective mass matrix that describes
unstable particles can be derived from propagator. This matrix automatically
respects the positivity and analyticity properties of the propagator.

In $QFT$, the physical masses are the poles of propagator $\Delta(z)$ or the zeroes of inverse propagator $\Delta^{-1}(z)$. Thus close to the poles, a linear approximation for $\Delta^{-1}$ should be
suitable,
\begin{equation}
\Delta^{-1}(z) \approx Az + B,
\end{equation}
where $A$ and $B$ are some constant matrices.

From the positivity of the propagator one can derive that  $A=A^\dag$ is a positive hermitian matrix.
If the property of positivity is true everywhere then $B=B^{\dagger}$.
In this case, the mass matrix is hermitian, its eigenvalues
are real and cannot describe unstable particles. However,
if one only wants to preserve this property in the upper (physical)
half plane $\Im(z) \geq 0$, it is enough to have $\Im(B) \geq 0$.
If this is so, then, writing $B= B_1 + i B_2, B_2 \geq 0$, one has

\begin{equation}
\Delta^{-1}(z) \approx \sqrt{A}\left( z + \frac{1}{\sqrt{A}} (B_1 + iB_2)
\frac{1}{\sqrt{A}}\right)\sqrt{A}
= \sqrt{A}\left(z - \left(M^{(2)}\right)\right)\sqrt{A}.
\end{equation}
To find the mass of the state we have to diagonalize matrix $M^{(2)}$. Thus matrix $M^{(2)}$
plays a role of a mass matrix. More accurately 

\begin{equation}
M^{(2)} \equiv (M- \frac{i}{2}\Gamma)^2.
\end{equation}
Thus mass matrix $M^{(2)}$ is defined in terms of propagator's matrices $A$ and $B$ 
\begin{equation}
M^{(2)}\equiv m^{(2)}-i\frac{{\Gamma^{(2)}}}{2} =
-\frac{1}{\sqrt{A}} (B_1 + iB_2)\frac{1}{\sqrt{A}}, {\rm with}\
{\Gamma^{(2)}} \geq 0, A=A^\dag.
\end{equation}
It is no longer hermitian and can  describe unstable kaons.
Since ${\Gamma^{(2)}} \geq 0$, the zeroes of the approximate inverse
propagator (poles of the approximate propagator) are located  in the lower (unphysical) half plane. 
The hermitian matrix $A$ normalizes the states.

  Near any given pole $z_i$ of the propagator we get some  \underline {new} mass 
matrix $M_i$. In other words for any given state we construct a \underline {new} Effective Hamiltonian.
Thus according to $QFT$ in the case of binary system like $(K^0, \bar{K}^0)$  we have to introduce  
\underline{two different} effective Hamiltonians - one for $K_S$ and another for $K_L$.

\subsection{$(K^0,\bar{K}^0)$ in QFT.}

Consider this construction in more details. 
For any momenta $z = q^2$ the propagator $\Delta(z)$  has two eigenvalues 
$\lambda_{\pm}(z)$ and four eigenvectors, i.e. two  (in) 
states $|R_{\pm}(z)>$ and two (out) states $<L_{\pm}(z)|$.
The same is true for the momenta near the pole 
$z_1=M_{L}^2$.  It is clear that one of the eigenvalues of  $\Delta^{-1}(z_1)$ has to be zero. Corresponding eigenvectors are the physical states that describe $K_L$ meson on-mass shell:
\begin{equation}
|R_{+}(z_1)>= |K_{L}>_{in}, \;\; <L_{+}(z_1)| = <K_{L}|_{out}.
\end{equation}  

Another eigenvalue of  $\Delta^{-1}(z_1)$ is  $2M_{L}^2$.
Corresponding eigenvectors
$|R_{-}(z_1)>$ and $ <L_{-}(z_1)| $ are non-physical spurious states, i.e they do not correspond to propagation of any particle on-mass shell. One can check that these states have zero norm.

Similar situation takes place for the momenta near the pole $z_2=M_{S}^2$. Thus we get four  on-mass shell states vs four spurious states. One can not delete spurious states since they make 
the system of eigenvectors complete. 

It is also clear that since we have different Hamiltonians for 
$K_L$ and $K_S$ the mixing parameters of eigenstates for two different Hamiltonians are also different. Thus $CPT$ symmetry
of fundamental $QFT$ does not entail that $CP$ parameter $\epsilon_{L}$ of $K_L$ is \underline{identical} to the one $\epsilon_{S}$ 
of $K_{S}$. That means that for $CPT$ invariant theory

\begin{equation}
|K_S>_{in}\sim (\left |\ K^0_1> +\epsilon_{S}^{in}|\ K^0_2>\right),\;\;
|K_L>_{in} \sim (\left |\ K^0_2> +\epsilon_{L}^{in}|\ K^0_1>\right).
\end{equation}
 and
 \begin{equation}
 \epsilon_L ^{in}\not= \epsilon_S^{in}
 \end{equation}
 
 This conclusion is in a sharp contrast with $QM$ treatment of $K$ mesons where $CPT$ symmetry and equality $\epsilon_L = \epsilon_S$ are just the same statement!  In the next section we shall
check whether this difference in formalism exhibits in physical observables.

\section {"Applications"}
The standard way to test $CPT$ symmetry is to measure semi-leptonic 
asymmetries $\delta_{L,S}$ and $A_{TCP}$ (see \cite{RPP}):
\begin{equation}
 \delta_{L,S}=\frac {|<\pi^{-} l^{+} \nu|K_{L,S}>|^2 -|<\pi^{+} l^{-} \nu|K_{L,S}>|^2}{|<\pi^{-} l^{+} \nu|K_{L,S}>|^2 + |<\pi^{+} l^{-} \nu|K_{L,S}>|^2};
 \end{equation}
and
 
\begin{equation}
 A_{TCP}=\frac {|<\pi^{-} l^{+} \nu|\bar{K}>|^2 -|<\pi^{+} l^{-} \nu|K>|^2}{|<\pi^{-} l^{+} \nu|\bar{K}>|^2 + |<\pi^{+} l^{-} \nu|K>|^2}.
 \end{equation}
 In the Standard Model $A_{TCP}$ can be also rewritten as

\begin{equation}
 A_{TCP}=\frac {|<\bar{K}(t_f)|\bar{K}(t_i)>|^2 -|<K(t_f)|K(t_i)>|^2}{|<\bar{K}(t_f)|\bar{K}(t_i)>|^2 +|<K(t_f)|K(t_i)>|^2}.
 \end{equation} 
  
  According to PDG booklet \cite{RPP} these two asymmetries are related to the difference of mixing parameters $\epsilon_{S,L}$:
  
 \begin{equation}
 A_{TCP}= \delta_S -\delta_L \simeq 2 \Re\; (\epsilon_S - \epsilon_L)
 \end{equation}
 
 Explicit perturbative calculation within our formalism gives
  \begin{equation}
 \delta_S = 2 \Re \;(\epsilon_S^{in}),  \delta_L = 2 \Re \;(\epsilon_L^{in}).
 \end{equation}
 Thus 
 \begin{equation}
\delta_S -\delta_L = 2 \Re\; (\epsilon_S - \epsilon_L)\not=0
 \end{equation}
for $CPT$ symmetric theory! 
 
The calculation of $A_{TCP}$ is slightly more subtle exercise (see \cite{0}).  We find that to construct
 correct perturbation theory for effective Hamiltonian one has to take into account spurious states. The contribution of these spurious states into $A_{TCP}$ exactly \underline{cancel} the contribution of physical states. As a result, 

\begin{equation}
A_{TCP}=0\not=\delta_S-\delta_L 
 \end{equation}
for $CPT$ symmetric theories. Thus $A_{TCP}$ is a \underline {good} test of $CPT$ violation.

\section {Numerical estimates}
 
In spite of conventional $QM$ treatment of $  ( K, \bar{K}) $ system mixing parameters for $K_L$ and $K_S$ states are different even for $CPT$ symmetric theory. Now we perform order of magnitude estimate of this difference $\epsilon_S^{in} - \epsilon_L^{in}$. 

To do that we need to know the dependence of $ (K,\bar{K})$  self-energy functions on momenta. The main contribution comes from  $K \leftrightarrow \bar{K}$ transition amplitude that takes place in the second order in weak interactions. To estimate the order of magnitude of the effect we calculate quark box diagram for $K \leftrightarrow \bar{K}$ transition. Inspecting this diagram we find \cite{0}
\begin{equation}
 \epsilon_S^{in} - \epsilon_L^{in} \sim \epsilon \frac{\Delta {m_{L,S}}}{m_K} \sim 10^{-17}.
 \end{equation}
 This effect is extremely small compared with the current experimental bounds on $CPT$ violation (see \cite {RPP, CPLEAR, KLOE}).

  As for theoretical estimates of expected $CPT$ violation effects they extend from the order of unity to zero.
At this conference Dolgov presented arguments that spin-statistics relation is different for neutrino. That will 
immediately break $CPT$ symmetry by order of unity.
To observe such violation of $CPT$ one can safely use conventional formalism from \cite {RPP}.

On the other hand if we believe in conventional field theory the only source of $CPT$ violation comes from non-locality of $QFT$ due to effects of gravity at small distances. This non-locality of $QFT$ should be very small due to Plank mass in the denominator. At best they are of the order $\sim {m_W}/{m_{Pl}} \sim 10^{-17}$. If one dreams to measure such effect one needs our formalism with all tiny corrections in order to separate genuine $CPT$ violation effects from the fake ones.  
 
\section{Conclusions}

Let me summarize the results.

 There is substantial difference between $QM$ and $QFT$ in treatment of binary systems. We find that correct formalism for $ (K,\bar{K})$ system imitates the effects that can be considered as  $CPT$  violation in conventional formalism.
Thus one has to remember that

{\bf {1)}} $QM$ is not appropriate framework for $CPT$ violation if effects are small;

{\bf {2)}} Asymmetry $\delta_{L}-\delta_{S}$ tests the difference
 $\epsilon_{L} - \epsilon_{S}$, not $CPT$ violation;

{\bf {3)}} Asymmetry $A_{TCP}$ measures $CPT$ violation.

\section {Acknowledgments}
 I would like to thank  the organizers of La Thuile conference, particular  Mario Greco,
 for their warm hospitality and for excellent conference.
This research was partly supported by RFBR
grant 05-02-17203.

\end{document}